\documentclass[pra,superscriptaddress,showpacs,twocolumn]{revtex4-1}

\usepackage{amsmath,amsfonts,amsthm,amssymb}
\usepackage{graphicx}
\usepackage{graphics}
\usepackage{braket}
\usepackage{setspace}
\usepackage{color}
\usepackage{ulem}
\usepackage{placeins}
\usepackage{xcolor}

\renewcommand{\emph}{\textit}

\usepackage{hyperref}

\begin{document}

\author{Marc~Fleury}

\title{Observations of Bell Inequality Violations with causal isolation between source and detectors}

\begin{abstract}
We report the experimental observations of Bell Inequality Violations (BIV) in entangled photons causally separated by a rotating mirror. A Foucault mirror gating geometry is used to causally isolate the entangled photon source and detectors.  We report an observed BIV of CHSH-$S = 2.30 \pm 0.07 > 2.00$.  This result rules out theories that explain correlations with traveling communication between source and detectors.   
\end{abstract}

%\pacs{03.65.Ud, 03.65.Wj, 06.20.Dk}

\maketitle

\section{Motivation and Hypothesis}
\subsection{On Bell Inequality Violations and Local Realism}
It is commonly accepted that the Aspect~\cite{Aspect1982} and Zeilinger~\cite{PhysRevLett.81.5039} experiments convincingly closed the locality loophole for Bell Inequality Violations (BIV) experiments. They do so by causally isolating the acts of measurements at two distinct points called Alice and Bob. Modern BIV experiments close the locality loophole in increasingly dramatic fashion. See ~\cite{Wengerowsky6684} for a submarine fiber implementation, ~\cite{Yin1140} for an orbital station implementation, and  ~\cite{ZeilingerStars} for measurement settings triggered by distant stars. The experiments all focus on isolating the measuring stations from one another: Alice and Bob are the ones causally separated, by land, sea, satellites and deep space.  While the experiments undoubtedly achieve observer to observer isolation we will argue in section~\ref{subsection:isolation} that these experiments do not isolate the emitting atoms from the observers. We will argue that in all these geometries, there is an ever present line of sight between the measuring apparatus and the source. We will motivate that such influence of the measuring stations on the source can account for BIVs in a classic local realistic ways in section~\ref{subsection:hypothesis}. It follows that we will lose BIV by removing the line of sight between source and detection. A Foucault mirror design gates and effectively removes the line of sight between observers and emitters. With this gate in place we perform a CHSH measure of BIV in a 2-channel setup.  In section~\ref{results} we report on the observation of CHSH $S=2.30>2$.  This rules out travelling influences between source and detectors, including super-luminal, as candidates for the Bell effect in section~\ref{conclusion}.

\subsection{On experimental locality: Observer to observer vs observer to emitting atoms}
\label{subsection:isolation}

\begin{figure}[htb]%
\centering%
\includegraphics[width=\columnwidth]{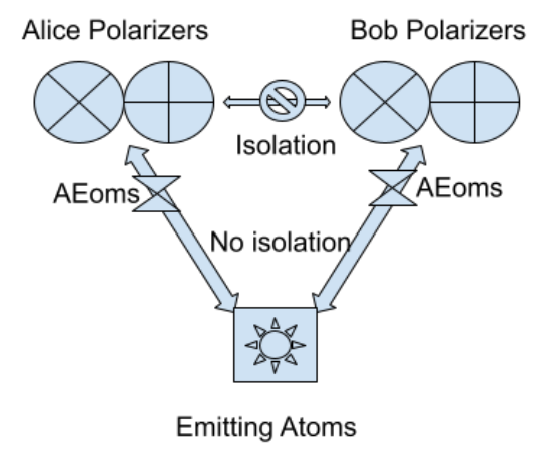}%
\caption{schematic representation of the isolation achieved in 4 channel Bell experiments with AOM and EOM. The isolation between Alice and Bob is complete. However there is no isolation between the four measures at Alice and Bob and the emitting atoms.}%
\label{fig:isolation}%
\end{figure}

We will now review the isolation experimentally achieved thus far in the current state of the art experiments. We will observe a static line of sight between emitting atoms and observing ones.  Bell famously prescribed "randomly setting the measure during the flight of the photons" as a way to close the locality loophole.  Modern experiments causally isolate the observers between themselves.  In the Aspect and Weihs-Zeilinger experiments, we never set the direction of the polarization analyzers during the flight of photons. This is in fact mechanically impossible~\cite{phdWeihs}. Instead, the analyzers are all preset and we dynamically route the photons to a randomly chosen analyzer.  This is achieved by randomly switching the path of the photons while in flight.  We then claim physical analogy to "setting the measures" but the measures were chosen more than they were set. The settings are in fact all preset, it is only the choice of the measure which is dynamic and random. To effect this random optical switch, the original experiments by A.A. used Acousto-Optic Modulators (AOMs) operating in the MHz domain~\cite{Aspect1982}, the modern state-of-the-art experiments in the vein of Weihs and Zeilinger, use Electro-Optic Modulators (EOMs) operating in the GHz domain~\cite{PhysRevLett.81.5039}~\cite{2015ZEIL}. These AEOM devices randomly switch the path of the photons and therefore randomly choose which one of the preset polarization analyzers will perform the actual measure.  Both experiments are based on a similar implementation concept of the optical switches and measures: they preset all polarization analyzers upfront and then choose one by randomly modulating the index of refraction of some media, either acoustically or electrically. In the EOM experiments, the emitting atoms inside the BBO crystal have a permanent line of sight to \emph{all} four polarizers with static settings. The EOM modulate the phases, the AOM modulate amplitudes.  The AOMs behave more like switching mirrors with a 4:1 reflection contrast. Therefore, a line of sight from the detectors to the emitting cascading atoms is also always present in the Aspect experiments. The presence of the AOM in the path does not remove this line of sight, it modulates the amplitudes of signals coming through it. State-of-the-art AOMs never realize a true zero in amplitude transmission, the residual signal is rather large. There is always a line of sight. 
Therefore we conclude that the atoms in the sources of both geometries (SPCD/BBO and cascading atoms) are connected to all four static measurements at all times.  The emission of the entangled photons happens with full visibility of the measurements at a fixed angle in both types of experiments. See fig. \ref{fig:isolation} for a schematic rendering of isolation achieved with these AEOMS. This line of sight was hiding in plain sight, so to speak. 

\subsection{Hypothesis: on a background influence between emitting atoms and measuring atoms}
\label{subsection:hypothesis}

With regards to Bell's theorem, such a background field would allow for Bell type correlations in simple local realistic ways. Such a de-localized background field trivially breaks the isolation assumption of the theorem according to ~\cite{Vervoort}. This is not in violation of, but rather in accord with, Bell's theorem. ~\cite{2016Schmelzers} considers ~\cite{Vervoort} wrong claiming it implies a super-luminal communication in the background. We will now briefly review the literature for BIVs and show that there is little agreement as to what is causing the Bell correlations.  For example, other modern local realist approach  hypothesize non-linear amplifications of the so-called zero-point field modes combined with a hypothetical threshold detection in Avalanche Photo Detectors (APDs)~\cite{LaCour}~\cite{Marshall}. Others, more generically, focus on Bohr contextuality pointing out that the measures always interfere with the measured, a point made clear since the Stern-Gerlach days. Both the photons in flight and the emitting source atoms are subject to dynamic influences~\cite{Contextuality}~\cite{Muchowski}. Others argue the physical fact that Bell type correlations emerge from classical Electro-Magnetism ~\cite{Ham}. Counting coincidences in a Bell game is after all measuring intensity at two different points. This intensity interferometer (as opposed to amplitude) is known as a Hanbury Brown-Twiss (HBT) interferometer~\cite{BrownTwiss}. Others still within the HBT interferometer approach, reconstruct the proper Bell intensity correlations with a classical EM local realist model ~\cite{Jung}. They hypothesize circularly polarized single photons coming out of the SPDC process in the BBO crystals. By definition of circular polarization. this introduces a quarter-wave phase delay between the horizontal and vertical modes of the BBO crystals. This is in clear opposition to the  understanding of BBO crystal emissions in ~\cite{phdWeihs}. There the horizontal and vertical modes are understood to be in phase which results in a "quantum superposition" of the linear modes. This is called a singlet or triplet state depending on the sign of the delay. In fact, as reported in Weihs' PhD thesis (German only) one fine-tunes the exit phases of the BBO modes so as to be in phase, a "sine-qua-non" condition of the observation of BIVs and a big part of the experimental approach.  This results in a 'quantum superposition' by erasing the which-path information contained in the phases.  Finally and as if to further confuse everyone in the ivory tower of B(e)aBel(l)~\cite{bell}, some recently report observations of Bell-type correlations using \emph{uncorrelated} laser sources.~\cite{Lasers}.  

\textbf{Hypothesis:} If a field is responsible for the Bell effect, then when we remove said field, we can anticipate a loss of Bell effect. (No line of sight? no BIV!). To test this background influence hypothesis and without further speculation as to the underlying physics involved, we now introduce a Foucault mirror designed to disrupt the EM line of sight between emitting atoms and measurement apparatus.  With this contraption we remove any EM background field between observers and emitters.  How does this affect a Bell-CHSH measure? We anticipate a possible loss of BVI.

\section{Experimental Setup}
\label{experiment}
\subsection{A 2-channel photon Entanglement measure using a rotating mirror}

\begin{figure}[htb]%
\centering%
\includegraphics[width=\columnwidth]{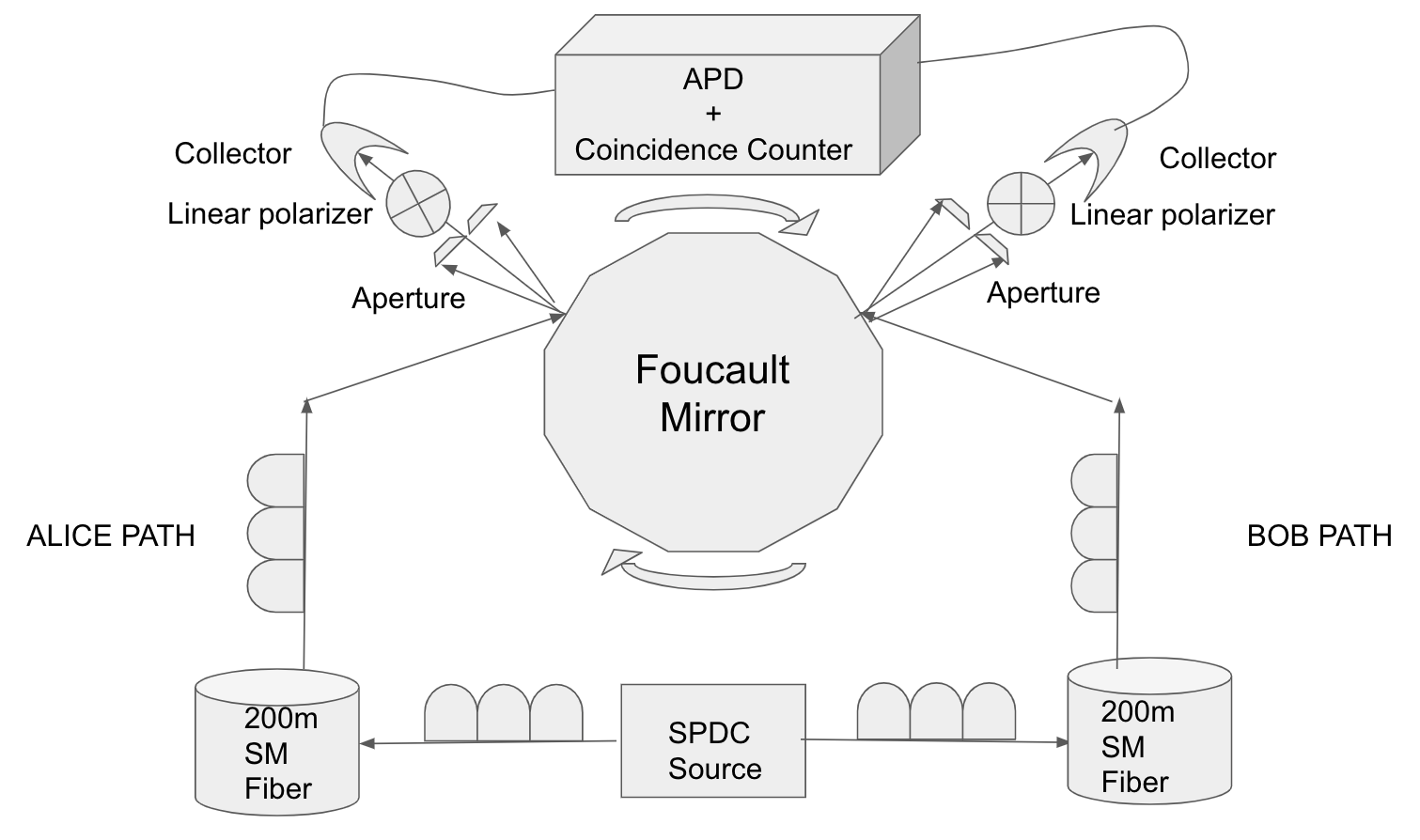}%
\caption{schematic layout.  SPDC source down in the middle. Alice and Bob photons are transported over 200 m Single-Mode Fiber. We compensate with 3 x bat ears. ($\tfrac{1}{4}$, $\tfrac{1}{2}$, $\tfrac{1}{4}$) wave delays both at the in and out of both fibers for a total of 4 bat ears.  Both paths are bounced off of the Foucault mirror which creates a line arc over an aperture at the detectors. We then filter through static computerized polarizers, into collectors and into the Avalanche Photo Detectors (APD) counters from qutools.}%
\label{fig:setup}%
\end{figure}

To test this hypothesis, we introduced a spinning Foucault mirror in a classic 2 channel Bell experiment.  See fig. \ref{fig:setup}. This is a table-top 2-channel experiment showing a Bell-CHSH Inequality Violation as opposed to the 4-channel detection loophole experiments reviewed above. We remove the line of sight by introducing a rotating Foucault mirror, inspired from the eponymous 1850 measure of the speed of light. This connects the components only for a brief amount of time and the source is optically isolated from the measures for the vast majority of the time. Contrast this to the AEOMs experiments, where the source is continuously connected to the detectors.  Here we only focus on the fiber communication, we do not care about physical line-of-flight isolation, in fact everything fits on an table top optical bread board. This type of causal isolation of the source is new and therefore, to the best of our knowledge, an experimental first. 

An entanglement source commercially available from quTools, GmbH,  allows for a classic 2-channel singlet experiment showing Bell-CHSH violations. A rotating mirror (\(1000\)Hz, \(34\) facets) purchased from Cambridge Technologies (model SA34 P/N:1-1-3304-001-00) completes the setup. The geometrical arc described by the beams reflected off each of the rotating mirror facets covers the slits over Alice and Bob. See fig. \ref{fig:setup}. The result is a gate: only when the rotating beams cover the openings over the slits are the gates opened and the components connected. 

\paragraph{\textbf{Time of aperture:}} We can adjust the time this window stays open and during which the paths are connected. The time of aperture is determined by the aperture width of the slits \(A=10^{-3}\)m, the rotational velocity \(w=10^{3}\)Hz, and the physical distance between the slits and the mirror on our bread board \(R=34\cdot 10^{-2}\)m: 
\begin{equation}
T_{on} = \frac{A}{2\pi R w} = 4.7\cdot 10^{-7}s
\end{equation} 
Since we use the same mirror for both paths, the shutter effects at both slits are in phase. The slits are alight at the same time and we thus detect coincidences in the APDs. Note that two different mirrors would not be in phase and we would not detect coincidences in the counters because the APDs would not be alight at the same time, or at least within the $20$ns coincidence window programmed in hardware.  

\paragraph{\textbf{Causal Isolation:}}
During the time of aperture, photons will travel a distance of $140\,$m. We introduce $200$m of single-mode fibers between source and mirror as in fig. \ref{fig:setup}. This achieves isolation of detectors and crystal when separated by the fibers.  In our table top setup, we only consider fiber isolation. The detectors are physically sitting next to each other on our bread board. They are within $10cm$ physical distance but their fiber optical paths are separated by $400$m ($2 \times 200$m) of single-mode transport. Detector to detector isolation is a byproduct of source to detector fiber isolation as we double the distance of fiber transport needed to $400$m.\newline

\paragraph{\textbf{Degradation:}}
Luminosity and time of aperture are proportional.  The shorter the time gate, the shorter the exposure to the light and the shorter the collection of photons. This is true of single counts.  The coincidence count is proportional to the single count, and not a quadratic of the single count. As we have observed, the mirror shines the separate individual slits at the same time, the gate effects at distant slits are in sync. This is the reason we detect coincidences in the first place. This yields a linear relationship between coincidences and singles, mainly due to the quantum efficiency of the detectors.  The floor in our experimental setup is set by the dark count of singles and the dark count of coincidences. With rotation, the photons are spread over $2\,$ lines that repeat $1000\,$ times per second. We calculate the signal degradation as: 
\begin{equation}
D_{th} = \frac{A\cdot N}{2\pi R} = 0.016
\end{equation} 
with the aperture size $A = 10^{-3}\,$m, the radius mirror to slit $ R =  0.34\,$m and the number of facets $N =34$.

\section{Results}
\label{results}

\paragraph{\textbf{Dark counts and detection window:}}
Our APDs have individual dark count rates of $1300\,$s$^{-1}$ and $600\,$s$^{-1}$, and the dark count coincidence rate is approx $300\,$hr$^{-1}$.
The time for the detection window of coincidences in APDs is $20\,$ns and is set in hardware by quTools.  We do not control this time window parameter. See tab. \ref{tab:results}.  \newline

\paragraph{\textbf{Degradation:}} The results of measurements for determining the luminosity and signal degradation due to the rotating Foucault mirror can be seen in tab. \ref{tab:results}. The dark counts of the detectors are measured in order to be subtracted from the results in further calculations. 
\begin{equation}
D_{ex} = \frac{\mathrm{Signal\;with\;rotation}}{\mathrm{Signal\;without\;rotation}}=0.018\;
\label{eq:}
\end{equation}
 The first observation is that we retain a clear coincidence signal, well above the noise level.   The degraded coincidence signal in our experiments is on the order of $30,000\,$hr$^{-1}$ and the noise for said coincidences is measured at $300\,$hr$^{-1}$.  We have a clear $100:1$ contrast in the signal-to-noise ratio. The ratio between singles and coincidences is due to the quantum efficiency of the detectors. The higher value of the experimentally observed signal degradation could be due to the non-linear quantum efficiency of the detectors: The higher the measured count rate, the lower the quantum efficiency.\newline

\begin{table}%
\caption{Results of signal luminosity and degradation measurements, in photons per second.}\vspace{0.5em}
\begin{tabular}{l|ccc}
 & Single 0 (/s) & Single 1 (/s) & Coinc. (/s) \\ \hline
Dark Counts & 1300 & 600 & 0.08 \\
No rotation & 33894 & 20329 & 389 \\
With rotation & 2301 & 1098 & 7 \\ \hline
Degradation $D_{ex}$ & $0.031\pm0.003$ & $0.025\pm0.004$ & $0.018\pm0.008$ \\
\end{tabular}
\label{tab:results}
\end{table}

\paragraph{\textbf{Bell CHSH observation:}}
We have observed BIV of Bell-CHSH $S = 2.30 \pm 0.07$. See tab. \ref{tab:chsh}.
 
\begin{table}%
\caption{CHSH inequality: observed. $S=2.302\pm0.071$.  There are 16 measurement settings for the polarizers. In blue the individual counts and in grey to the right the accidental coincidence counts (noise because of single counts). For example, the first cell is 0:22.5, the coincidence count is 226 photons and the noise is 5 photons.  The integration time is 60,000ms (1 minute per measure, 16 measures, 16 minutes total). The 1-minute per-measure setting is the maximum setting we can program within the quTools environment. } \vspace{0.5em}
\begin{tabular}{r|rrrr}
Angle in $^\circ$ & 0 & 45 & 90 & 135 \\ \hline
22.5 & \textcolor{blue}{226}\textcolor{gray}{-5} & \textcolor{blue}{85}\textcolor{gray}{-4} & \textcolor{blue}{42}\textcolor{gray}{-4} & \textcolor{blue}{184}\textcolor{gray}{-4} \\
67.5 & \textcolor{blue}{70}\textcolor{gray}{-5} & \textcolor{blue}{34}\textcolor{gray}{-4} & \textcolor{blue}{182}\textcolor{gray}{-4} & \textcolor{blue}{239}\textcolor{gray}{-5} \\
112.5 & \textcolor{blue}{46}\textcolor{gray}{-5} & \textcolor{blue}{187}\textcolor{gray}{-5} & \textcolor{blue}{227}\textcolor{gray}{-4} & \textcolor{blue}{70}\textcolor{gray}{-4} \\
157.5 & \textcolor{blue}{198}\textcolor{gray}{-5} & \textcolor{blue}{217}\textcolor{gray}{-5} & \textcolor{blue}{88}\textcolor{gray}{-4} & \textcolor{blue}{34}\textcolor{gray}{-4} \\
\end{tabular}
\label{tab:chsh}
\end{table}

\section{Conclusion}
\label{conclusion}
   We demonstrated that the AEOMs Bell experiments close the locality loopholes between observers but do not causally isolate the sources from the observers in classic 4-channel loophole experiments. We introduced a spinning mirror to create such isolation in a 2-channel Bell experiment. By intermittently removing this hypothetical background we anticipated a loss of the Bell effect (loss of BVI). We reported instead on observed Bell violations.  \newline
   Regarding a hypothetical background mediating an influence from the measurement apparatus to the emitting atoms and which is responsible for the Bell effect; we will prove that the influence cannot be a traveling wave. To prove this statement, we first observe that the time the gate is open is $4.7\cdot 10^{-7}$s and that the distance covered by photons in the single-mode fibers during this time is approximately $140$m.  Let us assume, for the sake of argument, that the influence from the detector to the source is instantaneous. The BBO crystal would start emitting under the influence of the detector as soon as the gate opens. The photons informed by this instantaneous influence would then still need to cover the $200$m of single-mode fiber that separates them from the detectors. By the time the photons reach the mirror, the slit is not in view anymore.  These photons are not detected.  We then observe that this is also true of a slower influence signal: the influence will simply take longer to reach the source and delay the return trip by as much.  We therefore conclude that the influence cannot be that of a travelling wave, not even super-luminal, not even an instantaneous wave. QED.  
   
\section{Discussion} 
For discussion we will offer the following logical derivation.  
Either: 
  \begin{itemize}
      \item 1/ There is no background influence, or
      \item 2/ There is a background influence, and either
      \begin{itemize}
          \item 2a/ it is a traveling local wave, or
          \item 2b/ not (standing wave or deBroglie hypothesis)
      \end{itemize}
  \end{itemize}
  
The logical structure of the above is a list enumeration of the possible scenarios:  $\nexists$ background $\lor (\exists$ background $\land ($ standing $\lor \neg $standing)). This is declarative, there are no physics yet. \newline

One can argue in scenario 1 that there is no background field it just does not exist or something else is responsible for the Bell effect. All we have done here is reduce luminosity and further validate orthodoxy. In scenario 2 we still consider the existence of a background influence. The conclusion of this paper was to experimentally rule out option 2a.   

Therefore we are left with option 2b.  The wave responsible for Bell is NOT a traveling wave. What form could this wave be? Standing waves come to mind. The author looks to the Hydrodynamic Walkers for visual inspiration ~\cite{Walkers}. A droplet is bouncing and creating impact waves.  Behind are standing waves. The sum of these standing waves is the wake. A chaotic form of time-averaged wake emerges ~\cite{budanur2018state}. This is such an example of memory we derive from the dynamic fields of the hydrodynamic walkers. These are the Faraday wave-fields which are manifested in the hydrodynamic walkers ~\cite{Bush2020}. This is also in alignment with the deBroglie's intuition of matter modelled with standing waves about singular points. In the Hydrodynamic Quantum Analog field (HQA) of the walkers, the walkers follow a cavity geometry which is represented by a ghostly but emerging mean field which includes a form of echo-location. This mean field of standing waves represents a memory of past events, the wake, and usually includes information from the enclosing cavity, in the form of standing waves of the cavity, and this resonant memory guides the particles. 

 \section{Acknowledgments}
 We wish to thank Emmanuel Fort, Herman Batelaan, Gregor Weihs, and Louis Vervoort for in depth feedback on this paper, Henning Weier and Nico Klein of QuTools for technical assistance. This work was self funded.
\bibliography{BellViolations}

\begin{thebibliography}{10}

\bibitem{Aspect1982}
Alain Aspect, Jean Dalibard, and Gérard Roger.
\newblock Experimental test of bell's inequalities using time-varying
  analyzers.
\newblock {\em Phys. Rev. Lett.}, 49:1804, 12 1982.

\bibitem{PhysRevLett.81.5039}
Gregor Weihs, Thomas Jennewein, Christoph Simon, Harald Weinfurter, and Anton
  Zeilinger.
\newblock Violation of bell's inequality under strict einstein locality
  conditions.
\newblock {\em Phys. Rev. Lett.}, 81:5039--5043, Dec 1998.

\bibitem{Wengerowsky6684}
S{\"o}ren Wengerowsky, Siddarth~Koduru Joshi, Fabian Steinlechner, Julien~R.
  Zichi, Sergiy~M. Dobrovolskiy, Ren{\'e} van~der Molen, Johannes W.~N. Los,
  Val Zwiller, Marijn A.~M. Versteegh, Alberto Mura, Davide Calonico, Massimo
  Inguscio, Hannes H{\"u}bel, Liu Bo, Thomas Scheidl, Anton Zeilinger,
  Andr{\'e} Xuereb, and Rupert Ursin.
\newblock Entanglement distribution over a 96-km-long submarine optical fiber.
\newblock {\em Proceedings of the National Academy of Sciences},
  116(14):6684--6688, 2019.

\bibitem{Yin1140}
Juan Yin, Yuan Cao, Yu-Huai Li, Sheng-Kai Liao, Liang Zhang, Ji-Gang Ren,
  Wen-Qi Cai, Wei-Yue Liu, Bo~Li, Hui Dai, Guang-Bing Li, Qi-Ming Lu, Yun-Hong
  Gong, Yu~Xu, Shuang-Lin Li, Feng-Zhi Li, Ya-Yun Yin, Zi-Qing Jiang, Ming Li,
  Jian-Jun Jia, Ge~Ren, Dong He, Yi-Lin Zhou, Xiao-Xiang Zhang, Na~Wang, Xiang
  Chang, Zhen-Cai Zhu, Nai-Le Liu, Yu-Ao Chen, Chao-Yang Lu, Rong Shu,
  Cheng-Zhi Peng, Jian-Yu Wang, and Jian-Wei Pan.
\newblock Satellite-based entanglement distribution over 1200 kilometers.
\newblock {\em Science}, 356(6343):1140--1144, 2017.

\bibitem{ZeilingerStars}
Dominik Rauch, Johannes Handsteiner, Armin Hochrainer, Jason Gallicchio,
  Andrew~S. Friedman, Calvin Leung, Bo~Liu, Lukas Bulla, Sebastian Ecker,
  Fabian Steinlechner, Rupert Ursin, Beili Hu, David Leon, Chris Benn, Adriano
  Ghedina, Massimo Cecconi, Alan~H. Guth, David~I. Kaiser, Thomas Scheidl, and
  Anton Zeilinger.
\newblock Cosmic bell test using random measurement settings from high-redshift
  quasars.
\newblock {\em Phys. Rev. Lett.}, 121:080403, Aug 2018.

\bibitem{phdWeihs}
Gregor Weihs.
\newblock {\em Ein Experiment zum Test der Bellschen Ungleichung unter
  Einsteinscher Lokalit}.
\newblock PhD thesis, Universitat Wien, 1999.
\newblock In german, available eletronically at
  www.uibk.ac.at/exphys/photonik/people/gwdiss.pdf.

\bibitem{2015ZEIL}
Marissa Giustina, Marijn~A. Versteegh, Sören Wengerowsky, Johannes
  Handsteiner, Armin Hochrainer, Kevin Phelan, Fabian Steinlechner, Johannes
  Kofler, Jan-Åke Larsson, Carlos Abellán, Waldimar Amaya, Valerio Pruneri,
  Morgan~W. Mitchell, Jörn Beyer, Thomas Gerrits, Adriana~E. Lita, Lynden~K.
  Shalm, Sae~Woo Nam, Thomas Scheidl, Rupert Ursin, Bernhard Wittmann, and
  Anton Zeilinger.
\newblock Significant-loophole-free test of bell’s theorem with entangled
  photons.
\newblock {\em Physical Review Letters}, 115(25), Dec 2015.

\bibitem{Vervoort}
Louis Vervoort.
\newblock No-go theorems face background-based theories for quantum mechanics.
\newblock {\em Foundations of Physics}, 46(4):458–472, Dec 2015.

\bibitem{2016Schmelzers}
I.~Schmelzer.
\newblock About a “nonlocal” local model considered by l. vervoort, and the
  necessity to distinguish locality from einstein locality.
\newblock {\em Foundations of Physics}, 47(1):113–116, Oct 2016.

\bibitem{LaCour}
Brian~R. La~Cour and Thomas~W. Yudichak.
\newblock Classical model of a delayed-choice quantum eraser.
\newblock {\em Phys. Rev. A}, 103:062213, Jun 2021.

\bibitem{Marshall}
Alberto Casado, Trevor~W. Marshall, and Emilio Santos.
\newblock Type ii parametric downconversion in the wigner-function formalism:
  entanglement and bell’s inequalities.
\newblock {\em Journal of the Optical Society of America B}, 15(5):1572, May
  1998.

\bibitem{Contextuality}
Th~Nieuwenhuizen.
\newblock Is the contextuality loophole fatal for the derivation of bell
  inequalities?
\newblock {\em Foundations of Physics}, 41:580--591, 03 2011.

\bibitem{Muchowski}
Eugen Muchowski.
\newblock On a contextual model refuting bell{\textquotesingle}s theorem.
\newblock {\em {EPL} (Europhysics Letters)}, 134(1):10004, apr 2021.

\bibitem{Ham}
B.~Ham.
\newblock The origin of anticorrelation for photon bunching on a beam splitter.
\newblock {\em Scientific Reports}, 10, 2020.

\bibitem{BrownTwiss}
R.~Hanbury {Brown} and R.~Q. {Twiss}.
\newblock {Correlation between Photons in two Coherent Beams of Light}.
\newblock {\em \nat}, 177(4497):27--29, January 1956.

\bibitem{Jung}
Kurt Jung.
\newblock Polarization correlation of entangled photons derived without using
  non-local interactions.
\newblock {\em Frontiers in Physics}, 8:170, 2020.

\bibitem{bell}
J.~S. Bell and Alain Aspect.
\newblock {\em The theory of local beables}, page 52–62.
\newblock Cambridge University Press, 2 edition, 2004.

\bibitem{Lasers}
M.~Iannuzzi, R.~Francini, R.~Messi, and D.~Moricciani.
\newblock Bell-type polarization experiment with pairs of uncorrelated optical
  photons.
\newblock {\em Physics Letters A}, 384(9):126200, Mar 2020.

\bibitem{Walkers}
Y.~Couder, S.~Protiere, E.~Fort, and A.~Boudaoud.
\newblock Walking and orbiting droplets.
\newblock {\em Nature}, 437:7056, Sep 2005.

\bibitem{budanur2018state}
Nazmi~Burak Budanur and Marc Fleury.
\newblock State space geometry of the chaotic pilot-wave hydrodynamics.
\newblock 2018.

\bibitem{Bush2020}
Bush~JWM Durey~M.
\newblock Hydrodynamic quantum field theory: The onset of particle motion and
  the form of the pilot wave.
\newblock {\em Front. Phys. 8:300.}, 2020.

\end{thebibliography}

\end{document}